\documentclass{entcs}
\usepackage{prentcsmacro}
\usepackage{graphicx}

\def\lastname{Abramsky}
\begin{document}

\begin{frontmatter}
  \title{What are the fundamental structures of concurrency? We still don't know!} 
  \author{Samson Abramsky \thanksref{ALL}\thanksref{myemail}}
  \address{Oxford University Computing Laboratory\\
    Oxford, U.K.} \thanks[ALL]{This research was supported by U.K. EPSRC} \thanks[myemail]{Email:
    \href{mailto:samson@comlab.ox.ac.uk} {\texttt{\normalshape
        samson@comlab.ox.ac.uk}}}
\begin{abstract} 
Process algebra has been successful in many ways; but we don't yet see the lineaments of a fundamental theory. Some fleeting glimpses are sought from Petri Nets, physics and geometry.
  \end{abstract}
\begin{keyword}
  Concurrency, process algebra, Petri nets, geometry, quantum information and computation.
\end{keyword}
\end{frontmatter}

\section{Process Calculi as Generic Theories}
What counts as a successful theory in Computer Science?
Consider obvious exemplars such as
\begin{itemize}
\item Process Calculi
\item Type Systems
\item Model-checking
\end{itemize}
It is not the case that there is a single agreed model,
notation, formalism, tool or language in any of the above areas. 
In fact there are a profusion of all of these, although some
have been particularly influential. 
(Insert your favourite examples here \ldots )

\paragraph{The `Next 700 $\cdots$' syndrome}
Is this profusion a `scandal' of our subject?
I used to think so --- and I wasn't alone (e.g. Robin Milner's quest to find the `$\lambda$-calculus of concurrency'). 
Now I am not so sure.

\paragraph{It's the \itshape{Paradigms}!}
The paradigms and tool-kits, both technical 
and conceptual, provided by these theories have been deeply absorbed
by the research communities and have increasingly influenced
applications.

\noindent \textbf{Examples:} 
\begin{itemize}
\item labelled
transition systems and bisimulation
\item naming and scope restriction and
extrusion
\item 
the automata-theoretic paradigm for model-checking
\item the
type systems paradigm, with compositional typing rules for
terms-in-context, and key structural properties such as Subject
Reduction. 
\end{itemize}

\paragraph{By their fruits shall ye know them.}
These tool-kits are \emph{the real fruits of these theories}. They may be
compared to the traditional tool-kits of physics and engineering:
Differential Equations, Laplace and Fourier Transforms, Numerical 
Linear Algebra, etc.

They can be applied to a wide range of situations, going well beyond those originally envisaged, e.g.
Security, Computational Biology, Quantum Computing, etc.
So, is everything in the garden rosy?

\paragraph{Dreams of Final Theories}
But can we do better than this?
After all, in physics there \emph{are} great theories which transcend
mere tool-kits.
We largely lack such theories, in Computer Science as a whole, and in concurrency and process calculus in particular.
Is this unavoidable, as part of the nature of our subject, or will such
theories emerge?

Some may find such questions uninteresting, or even meaningless; they can safely stop reading here.

\section{Process Calculi vs. Concurrency Theory}

1980 marked the start of a new era in concurrency theory, but not its beginning. A meaningful theory of concurrency, incorporating some profound insights, had been originated by Petri in the 1960's, and Net theory, as well as other approaches to concurrency, continues to be actively developed.

There is no doubt that the advent of algebraic process calculi marked a decisive advance in concurrency theory, in particular in the use of compositional algebraic methods for the description of complex systems. It is often the case, though, that when an advance is made, something valuable is also lost, or at least, temporarily forgotten.

Let us start with the problem of \emph{canonicity} --- the `next 700 process algebras' syndrome. In a sense, the very success of the paradigmatic tool-kit, as described in the previous section, is also the source of the problem. It is too easy to cook up yet another variant process calculus or algebra; there are too few constraints. This \emph{plasticity of definitions} has become so familiar in our field that we may not be aware of it as an issue. The mathematician Andr\'e Weil apparently compared finding the right definitions in algebraic number theory --- which was like carving adamantine rock --- to making definitions in the theory of uniform spaces (which he founded), which was like sculpting with snow. In concurrency theory, we are very much at the snow-sculpture end of the spectrum. We lack the kind of external reality, whether it comes from fundamental mathematical objects like the integers, or manifolds, or differential equations, or from physical reality as determined by experiment, which is hard and obdurate, and resistant to our definitions. Is this a necessary feature of our existence, or have we just not yet found the real bedrock?

An important quality of Petri's conception of concurrency is that it \emph{does} seek to determine fundamental concepts: causality, concurrency, process, etc. in a syntax-independent fashion. Another important point, which may originally have seemed merely eccentric, but now looks rather ahead of its time, is the extent to which Petri's thinking was explicitly influenced by physics (see e.g. \cite{Pet}. As one example, note that K-density comes from one of Carnap's axiomatizations of relativity).  To a large extent, and by design, Net Theory can be seen as a kind of \emph{discrete physics}: \textbf{lines} are time-like causal flows, \textbf{cuts} are space-like regions, \textbf{process unfoldings} of a \textbf{marked net} are like the solution trajectories of a differential equation. 

This acquires new significance today, when the consequences of the idea that `Information is physical' are being explored in the rapidly developing field of quantum informatics. Moreover, the need to recognize the spatial structure of distributed systems has become apparent, and is made explicit in formalisms such as the Ambient calculus, and Milner's bigraphs.

\subsection*{Some morals}
\begin{itemize}
\item The genius, the success, and the limitation of process calculi is their \emph{linguistic character}. This provides an ingenious way of studying processes, information flow, etc. without quite knowing, independently of the particular linguistic setting, what any of these notions are. One could try to say that they are implicitly defined by the calculus. But then the fact that there are so many calculi, potential and actual, does not leave us on very firm ground.

We lack syntax-independent, \emph{intrinsic} definitions of the fundamental notions of concurrency theory. Net theory and some related approaches (e.g. event structures) still offer the best extant accounts of these issues. But we are still far from home.

Thus for example consider the issue of \emph{expressiveness}. There are some fragmentary results, but there is no single compelling notion of `expressive completeness' for a process calculus, or of a `Church's thesis for concurrency'.

\item We must now also acknowledge that we do not have sole ownership of the notions of information, process, etc. Physics and biology are also interested --- and they are at our gates! This presents us with a challenge, and perhaps also an opportunity for some new thinking on these issues.
\end{itemize}

\section{New directions: biology, physics or geometry?}

A major recent development in process calculi has been their application to biological modelling. This represents perhaps the first substantial example of a trend which, in my view, will form a major part of the future development of our subject: the spreading outwards of ideas developed in Computer Science, of the tool-kits we discussed in Section~1, to other scientific disciplines.
Provided there is a real engagement between the  CS bio-concurrency community and the biologists, this development has great promise.

However, while biological modelling will surely make new demands on process calculi, and hence lead to new developments (the next 700 biological process calculi?), I don't believe it is likely to lead to foundational advances for the issues we are discussing. Biology's foundational and conceptual structures are, if anything, much more plastic than those of Computer Science --- for which, of course, it compensates by the exuberant richness and the sheer concrete reality of the existence proofs which it studies.

There is, perhaps, more prospect for guidance in finding fundamental notions of process, information flow, etc. from the rapidly developing interface between Computer Science and Physics, which has grown up around quantum informatics. We have already discussed how Petri's development of Net theory was influenced by ideas from physics, and indeed provides some of the ingredients of a discrete physics. (One feature conspicuously \emph{lacking} there is an account of the non-local information flows arising from entangled states, which play a key role in quantum informatics. Locality is so plausible to us --- and yet, at a fundamental physical level, apparently so wrong!). Meanwhile, there are now some matching developments on the physics side, and a greatly increased interest in discrete models. As one example, the causal sets approach to discrete spacetime of Sorkin et al. \cite{Sor} is very close in spirit to event structures.

My own recent work with Bob Coecke on a categorical axiomatics for Quantum Mechanics \cite{AbrCoe2,AC1.5}, adequate for modelling and reasoning about quantum information and computation, is strikingly close in the formal structures used to my earlier work on Interaction Categories \cite{InteractionCats} --- which represented an attempt to find a more intrinsic, syntax-free formulation of concurrency theory; and on Geometry of Interaction \cite{Abr}, which can be seen as capturing a notion of interactive behaviour, in a  mathematically rather robust form, which can be used to model the dynamics of logical proof theory and functional computation.

This work admits a striking (and very useful) diagrammatic presentation, which suggests a link to geometry --- and indeed there are solid connections with some of the central ideas relating geometry and physics which have been so prominent in the mathematics of the past 20 years.\footnote{For the afficionado: the diagrammatics of our categories connect with categorical approaches to the Jones polynomial and other topological invariants, which in turn are strongly connected to quantum groups and topological quantum field theories.}
We note also that, in a rather different style, the geometry of concurrency has been developed by Eric Goubault \cite{Gou} and others.
So, geometry may yet have an important role to play in concurrency theory.

\subsection*{Whither process calculus?}
If anything like these speculations comes to pass, I think process calculus will be raised to a new level. It will, perhaps, become truly \emph{the} calculus of a fundamental science of information dynamics.

\bibliographystyle{entcs}
\end{document}